\documentclass[aps,prb,reprint,groupedaddress]{revtex4-2}
\usepackage{graphicx}
\usepackage{bm}
\usepackage{xcolor}

\begin{document}

\newcommand{\be}{\begin{equation}}
\newcommand{\ee}[1]{\label{#1}\end{equation}}
\newcommand{\bem}{\begin{eqnarray}}
\newcommand{\eem}[1]{\label{#1}\end{eqnarray}}
\newcommand{\eq}[1]{Eq.~(\ref{#1})}
\newcommand{\Eq}[1]{Equation~(\ref{#1})}

\newcommand{\rc}[1]{\textcolor{red}{#1}}


\title{Reply to Comment on ``Ballistic SNS sandwich as a Josephson junction''}

\author{Edouard  B. Sonin}
\email[]{sonin@cc.huji.ac.il}

\affiliation{Racah Institute of Physics, Hebrew University of Jerusalem, Givat Ram, Jerusalem 9190401, Israel}


\date{\today}

\begin{abstract}
This is the Reply to the Comment  by E. Thuneberg refuting the paper by E. B.Sonin [Phys. Rev. B {\bf 104}, 094517 (2021)] as incorrect. The criticism is based on a misunderstanding of the  goal and  the approach of the paper and
does not provide any reasonable argument that it is incorrect.\end{abstract}


\maketitle

Before discussing the Comment by \citet{Thun} (referenced later as the Comment) I describe the problem addressed   in the paper  \cite{Son21,*Son21er} (further called the Paper). The goal was to calculate the current-phase curve of the long ballistic SNS sandwich by the method of the self-consistent field \cite{deGen}. In this method  an effective pairing potential is introduced, which transforms the second-quantization Hamiltonian with an electron interaction term into an effective Hamiltonian, which is quadratic in creation and annihilation electron operators. The effective Hamiltonian   can be diagonalized by the Bogolyubov--Valatin transformation. The parameters of the transformation are the two-component wave functions, which are solutions of the Bogolyubov--de Gennes equations.  The pairing potential in the effective Hamiltonian must satisfy the integral self-consistency equation. In the previous investigations \cite{Kul,Ishi,Bezug,Bard} and in the Paper this step was skipped. Instead a simple profile of the pairing potential was postulated:   the pairing potential (gap)  of  the constant modulus  $\Delta_0$  in superconducting layers, and zero pairing potential  in the normal layer. Then the problem reduces to the analytical solution of the Bogolyubov--de Gennes equations and subsequent summations over all quasiparticle states. Eventually one obtains the analytical solution of the problem, which is {\em exact} in the limit of small ratios of the gap to the Fermi energy and of the coherence length to the thickness $L$ of the normal layer.

 The Comment argues that my analysis contradicts ``a vast literature on the topic'' \cite{Kul,Ishi,Bezug,Bard}. In fact, the Paper confirmed  the current-phase curve obtained in previous calculations for multidimensional (2D and 3D) systems for any temperature and for the 1D system  for zero temperature.   Different results were obtained only for the 1D case at  high temperatures   (high compared with the Andreev level energy spacing but still low compared to the superconducting gap).  But the Paper presented a new approach to the problem and another physical picture of the phenomenon. Now  I explain why I was not satisfied  by ``a vast literature on the topic'':
 \begin{enumerate}

\item The model with the effective Hamiltonian is not gauge invariant. Therefore, there are solutions of the model violating the charge conservation law. They are mathematically correct, but are unphysical and should be filtered out. I was not satisfied with how it was done in the previous literature.

\item
The effect of parity of Andreev levels (odd versus even number of states) was not considered or even mentioned. The effect can be essential for the 1D case (in analogy with effect of parity of electron numbers in normal 1D rings). 

\item
There was no clarity about the effect of continuum states on the current in the normal layer, despite statements to the contrary by the  Comment. \citet{Ishi} argued that this effect was important. This view was supported in a number of later publications and repeated in the  Comment. But none of them compared contributions to the current from bound Andreev and continuum states. Moreover, \citet{Bard} reproduced the result of  \citet{Ishi} without taking into  account continuum states.

\end{enumerate}

Dealing with Problem $\# 1 $  the Paper suggested a remedy for the absence of the conservation law. Instead of the strict law, the  Paper imposed a softer condition that at least the total currents deep in all layers  are the same. The condition  can  be satisfied taking into account three contributions to the total current $J=J_s+J_v+J_q$: (i) The current $J_s$ induced by the phase gradient in the superconducting layers. In the Paper it was called  the Cooper-pair condensate, or simply the  condensate current.  (ii) The current $J_v$, which can flow  in the normal layer even if the Cooper-pair condensate is at rest  and all Andreev states are empty. In the Paper it was called the  vacuum current.  (iii) The current $J_q$  induced by nonzero occupation of Andreev states, i.e., by creation of quasiparticles. It was called the excitation current. The condensate motion produces the same current $J_s$   in superconducting and normal layers of the SNS sandwich, while the  vacuum  and  excitation currents  exist only in the normal layer. Thus, the charge conservation law requires that  the sum  of the  vacuum  and   the excitation currents $J_v+J_q$  always vanishes.

\begin{figure}[!t] 
\includegraphics[width=0.34 \textwidth]{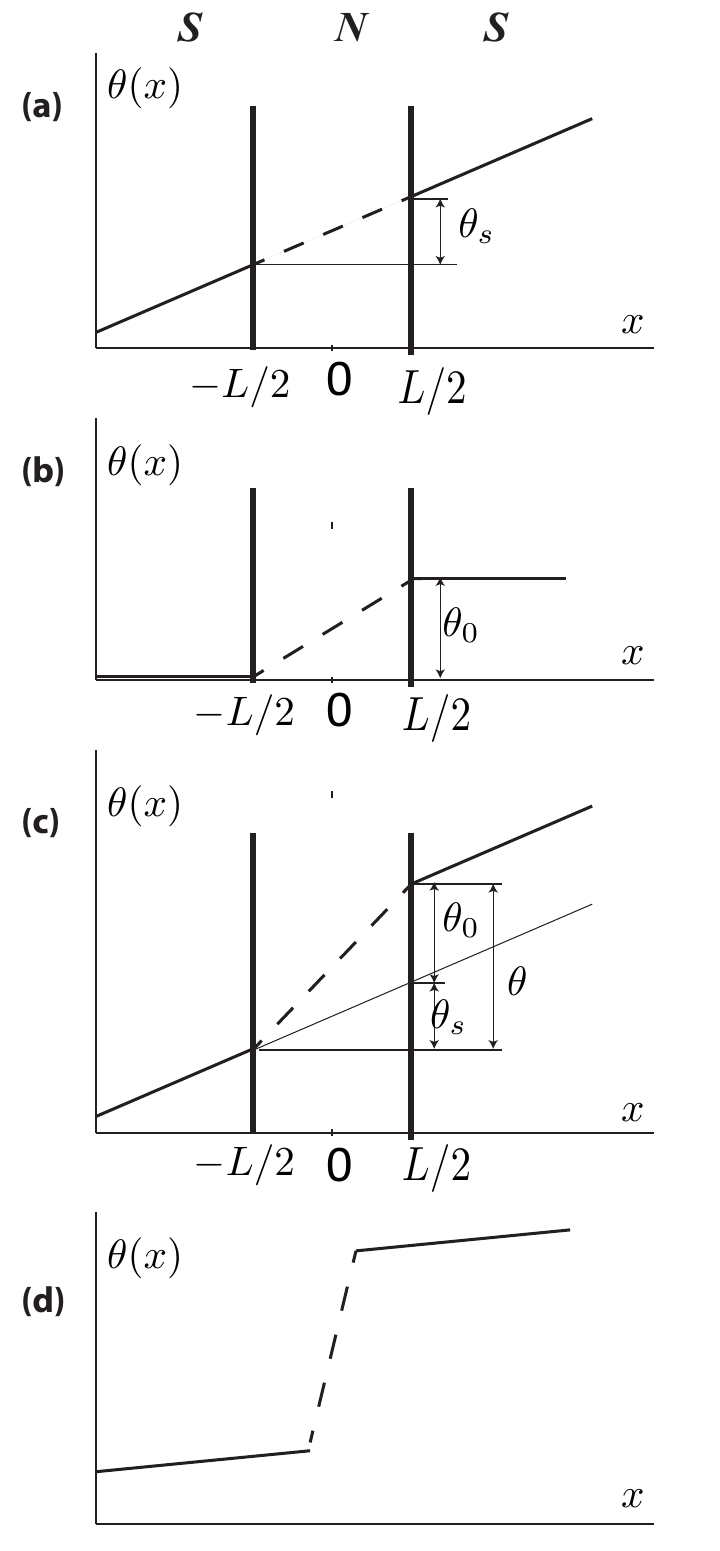} 
 \caption{The phase variation across the SNS sandwich.  (a) The condensate current produced by the phase gradient  $\nabla \theta$ in the superconducting layers.   In all layers the electric current is equal to $env_s$ where $n$ is the electron density and $v_s ={\hbar \over 2m}\nabla \theta$  is the superfluid velocity.  The phase $\theta_s =L\nabla \theta$, which is called the superfluid phase.  (b) The vacuum current produced by the phase $\theta_0$ called vacuum phase. The current is confined to the normal layer, there is no current in superconducting layers.  (c) The  superposition of the condensate  and the vacuum current. (d) The phase variation across a weak link. \label{f1}}
 \end{figure}

The phase variation  in states with the condensate and vacuum currents is shown  in Fig.~\ref{f1}. The condensate current $J_s=env_s$ [Fig.~\ref{f1}(a)] appears if there is the phase gradient $\nabla \theta$  in superconducting leads, which determines the superfluid velocity $v_s ={\hbar \over 2m}\nabla \theta$ ($n$ is the electron density).
The same condensate current in the normal layer requires the phase difference $\theta_s =L\nabla \theta$ across the normal layer called the superfluid phase.
 This is directly confirmed by the solution of the Bogolyubov--de Gennes equations. The vacuum current is determined by the  vacuum  phase $\theta_0$  [Fig.~\ref{f1}(b)].           
Figure~\ref{f1}(c) shows the phase variation at the coexistence of the condensate and the vacuum current. The total phase difference across the normal layer is the Josephson phase $\theta=\theta_s+\theta_0$. The phase profiles in the normal layer are shown in Fig.~\ref{f1} by dashed lines since this phase is not and cannot be determined because it is a phase of the order parameter $\Delta$, which vanishes  in the normal layer. Only the total phase difference across the normal layer appears in the Bogolyubov--de Gennes equations. Dashed lines simply show what the phase gradient would be if a normal metal were replaced by a superconductor.

 The previous literature \cite{Kul,Ishi,Bezug,Bard} and  the Comment considered the states without phase gradients  in superconducting layers. Thus, they  considered the case shown in Fig.~\ref{f1}(b) when  the current is determined by the vacuum phase $\theta_0$.

Ignoring the effect of the phase gradient in superconducting leads on the Josephson current was common in the past. By default it was supposed that this is not an issue because the gradients in leads are very  small. This is true for a weak link inside which the phase varies much faster than in leads as shown in Fig.~\ref{f1}(d).  As argued in the Paper,  the long SNS junction is not a weak link, and the phase gradient in leads does affect the current in the normal layer. Thus, one should determine   currents in normal and  superconducting layers  self-consistently. 
 
 The Comment denies the very existence of the problem with the charge conservation law   (Problem $\# 1 $) for vacuum current. The Comment discusses  the vacuum current  [see Fig.~1(c) in the Comment] not for 1D leads as in the Paper but for multidimensional leads. The goal of this ``trick'' (by the definition of the Comment itself) is to show that the vacuum current can flow without violation of the conservation  law. This 
sweeps the problem  under the carpet. A problem revealed in one case is denied because it does not exist  in another case. The logics of such an argument looks strange. The reason why the conservation law problem is absent (or, more carefully, is not so important) for multidimensional leads is clear: with multidimensional leads  the junction becomes a weak link  (see the  previous paragraph).

 The Comment does not see any  difference between  the condensate and the vacuum currents. The Paper discusses this difference in details. Tuning of  the phase $\theta_s$ producing the condensate current shifts  Andreev levels together with the gap edges, so that their relative positions do not vary  [see Fig.~4(b) in the Paper]. But at tuning  the phase $\theta_0$ connected with the vacuum current  Andreev levels move with the respect to gap edges and can cross them [see Fig.~4(a) in the Paper]. It is interesting that this phenomenon is mentioned in the Comment. I quote:  ''…the levels are shifted relative to the gap edge, and some new discrete levels may appear and some others disappear.'' But the Comment failed to notice that these processes are not possible at tuning of the phase  $\theta_s$. This justifies an introduction of two phases in addition to their sum (the Josephson phase) $\theta=\theta_0+\theta_s$, which is eventually present in the final current-phase relation.

\begin{figure}[!t] 
\includegraphics[width=0.4 \textwidth]{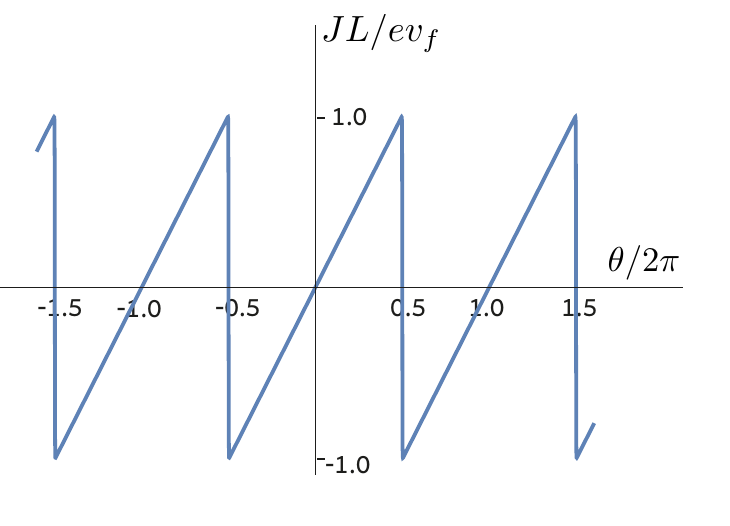} 
 \caption{The saw-tooth current-phase curve at zero temperature. \label{f2}}
 \end{figure}

Introduction of two phases $\theta_0$ and $\theta_s$ essentially revised the physical picture of charge transport through the ballistic SNS junction.   
Let us consider the current-phase curve at zero temperature (Fig.~\ref{f2}), which was the same in the previous literature \cite{Kul,Ishi,Bezug,Bard} and in the Paper. 
But, as mentioned above,  the previous literature and  the Comment considered the states without phase gradients  in superconducting layers, i.e.,  they  calculated the vacuum current as a function of the vacuum phase $\theta_0$ at $\theta_s=0$ [the case shown in Fig.~\ref{f1}(b)].  Meanwhile, the charge conservation law does not allow the vacuum current without the excitation current compensating it. At sloped segments of the $T=0$ current-phase curve there are no quasiparticles in Andreev states because all their energies are positive. Thus,   complicated  calculations of the vacuum current in the previous literature are not relevant for sloped segments of the current-phase curve. According to the Paper, at sloped segments  only the condensate current determined by the phase $\theta_s$ flows without violation of charge conservation law. The condensate  current   is simply determined from the principle of Galilean invariance valid at Andreev reflection at interfaces between layers despite the absence of translational invariance. This principle was formulated by \citet{Bard}, but they applied  it only to the normal layer ignoring that this principle requires that the same current flows also in superconducting layers. The vacuum current at zero temperature appears only at vertical segments of the $T=0$ current-phase curve at $\theta =\pi (2s+1)$ ($s$ is an integer) when  the energy of the  lowest   Andreev state reaches zero and its  occupation becomes  possible. This allows one to satisfy the condition that the sum of the vacuum and the excitation current must vanish. At $\theta \neq \pi (2s+1)$ (slope  segments) the charge transport through the junction does not differ from that in a uniform superconductor.

According to  the Comment, the Paper is incorrect because it ignored the contribution of continuum states to the current through the SNS junction \footnote{Here we discuss only the contribution of continuum states to the vacuum current, which flows in the normal layer while in superconducting leads the electron fluid is at rest. Continuum states give the main contribution to the condensate current, which flows in all layers without breaking the charge conservation law.    }. Formulating Problem $\# 3$  I mentioned that the statement of \citet{Ishi} about importance of this contribution was nothing more than a declaration, which was not supported by any calculation or estimation. Moreover, the end of the second paragraph of the Comment  warned against calculating contributions to the current from bound Andreev states and  continuum states separately. Thus, it is not only unknown from the Comment, which contribution to the current is more important, but also the Comment does not recommend  checking it. Without knowing how large the current in continuum states is there are no grounds for judging whether I was  correct or not ignoring the current in continuum states.

Meanwhile, there is no problem to receive an  exact expression for the contribution of continuum states to the vacuum current in the normal layer since the wave functions  of continuum states satisfying the Bogolyubov--de Gennes equations are known:
\be
J_{vC} ={e\over \pi\hbar}\int_\Delta^\infty  [{\cal T}(- \theta_0)-{\cal T}( \theta_0) ] d\xi,   
     \ee{cJ}
where
\bem
{\cal T}(\theta_0)= \frac{2\xi^2} {2\xi^2+\Delta_0^2\left[1-\cos\left({2\varepsilon m  L\over \hbar^2 k_f}-\theta_0\right)\right]}
 \nonumber \\
 = \frac{2\xi^2} {2\xi^2+\Delta_0^2\left\{1-\cos\left[{2(\varepsilon-\Delta_0) m  L\over \hbar^2 k_f}+2\pi \alpha-\theta_0\right]\right\}}
     \eem{Ta} 
is the transmission probability \footnote{The parameters of continuum scattering states were calculated by \citet{Bard} and reproduced  in the Paper. They are determined by  Andreev reflection at interfaces between layers. But \citet{Bard} did  not estimate the current in continuum states. This was done in the Paper. }  for a quasiparticle with the energy $\varepsilon$, and $\xi=\sqrt{\varepsilon^2-\Delta_0^2}$. 
The incommensurability parameter $\alpha$ is a  fractional part of the ratio of the gap to the Andreev interlevel spacing:
\be
{\Delta_0  L\over \pi \hbar v_f}=s_m +\alpha.
     \ee{inc}
An integer $s_m$ is chosen so that $0 <\alpha <1 $. In multidimensional cases integration over wave vectors transverse to the current  requires averaging of $J_{vC}$ over $\alpha$. After this  two terms in \eq{cJ} become independent from $\theta$, and the current  $J_{vC}$ totally vanishes. This refutes the claim  of  the Comment that the Paper erroneously ignored the contribution of continuum states to the current at least for multidimensional systems.

In the 1D case one should not average over $\alpha$, and the current $J_{vC}$ does not vanish. After some crude estimations I concluded in the Paper that nevertheless  in the limit $L\to \infty$ this current can be ignored also in the 1D case.  A more quantitative calculation would be useful, and the work in this direction is in progress.

Apparently the most nontrivial prediction of the Paper was the conclusion  about possibility of the anomalous current-phase relation with the ground state at nonzero phase $\theta$.
In the Paper the junction with such an anomaly was called thhe $\theta$ junction. But later  I became aware that earlier these junctions were known as $\varphi_0$ junctions \cite{Buzdin}.
The anomaly can be explained by simple arguments. In the ground state there is no condensate current ($\theta_s=0$), and it is sufficient to investigate the  dependence vacuum current versus phase $\theta_0$ [the case shown in Fig.~\ref{f1}(b)]. Definitely there  is an energy extremum at $\theta_0=0$, but one should check whether  it is a minimum or maximum.
The anomaly appears if the extremum is a maximum and  the second derivative of the energy, i.e, the first derivative of the current  ${dJ_v\over d\theta_0}+{dJ_v\over d\theta_q}$
 with respect to $\theta_0$  is negative. The derivative of the excitation current ${dJ_q\over d\theta_0}$ is always negative, and according to the Paper  at high temperatures its absolute value exceeds the positive derivative ${dJ_v\over d\theta_0}$. Thus, the energy extremum at $\theta_0=0$ is a maximum. This means that the total current decreases with growing small  phase, as we show now.
 
 The condition that $J_v+J_q$  vanishes must be checked taking into account that  the excitation current is determined by the total phase $\theta=\theta_0+\theta_s$:
 \be
 J_v+J_q={dJ_v\over d\theta_0}\theta _0+{dJ_q\over d\theta_0}(\theta _0+\theta_s)=0.
      \ee{}
 Then the total current  at small $\theta=\theta _0+\theta_s$ is
 \be
 J=J_s=env_s= {ev_f\over \pi L}\theta_s= {ev_f\over \pi L}\frac{{dJ_v\over d\theta_0}+{dJ_q\over d\theta_0}}{{dJ_v\over d\theta_0}}\theta.
         \ee{}
The sign of the current at growing positive $\theta$ is determined by the sign of ${dJ_v\over d\theta_0}+{dJ_v\over d\theta_q}$ indeed.

In the past $\varphi_0$ junctions ($\pi$ junction with $\varphi_0=\pi$ is the  most known example) usually were explained by magnetism of the normal  layer  \cite{Buzdin}. In our case there is no magnetism. This is an example of spontaneously broken time-reversal symmetry: if a $\varphi_0$ junction is put into a superconducting ring, a persistent current and a related magnetic  moment     appear in the ground state of the ring.

According to the Comment the most reliable method to solve the problem is the quasiclassical Green's function formalism but not the approach chosen in the Paper. I  fully respect  the Green's function formalism, which is useful for problems with interaction or disorder solved by the perturbation theory. However, we deal with the case without disorder or interaction. The problem is reduced to quadratures. There is an analytical solution  expressed in   {\em ab initio} sums and integrals, which, however, are not so simple for calculation. The Green's function formalism suggests a complicated chain of transformations dealing not with one but with a family of Green's functions. In the end after a number of assumptions one ends with other sums and integrals, which also require some efforts for calculations. I do not understand why this approach is more reliable than direct calculations of  {\em ab initio} sums and integrals.
 
The Paper presented a {\em quantitative} calculation of {\em ab initio} sums and integrals. One might expect that for a claim that a calculation is wrong some {\em quantitative} arguments were  necessary—namely, a calculation or at least a crude estimation disproving some step or assumption of the original calculation. One cannot find a single quantitative argument in the Comment. The only explanation why the calculation is incorrect is that the Paper ignored the contribution of continuum states. This is together with the statement that this contribution is unknown and does not deserve calculation
separately from the contribution of bound states.

In summary, the Comment does not contain anything that puts in doubt the method and the conclusions of the Paper. No evidence that they are incorrect was presented. That said,  I thank Erkki Thuneberg  for his the Comment and discussions, which urged me to  treat the continuum states more carefully. 


%


\end{document}